# Tunable tunnel barriers in a semiconductor via ionization of individual atoms

*Sara M. Mueller[1], Dongjoon Kim[2], Stephen R. McMillan[3], Steven J. Tjung[1], Jacob J. Repicky[1], Stephen Gant[1], Evan Lang[1], Fedor Bergmann[4], Kevin Werner[1], Enam Chowdhury[1], Aravind Asthagiri[2], Michael E. Flatté[3], and Jay A. Gupta[1*]*

[1]Department of Physics, Ohio State University, Columbus, OH 43210

[2]Department of Chemical and Biomolecular Engineering, Ohio State University, Columbus, OH 43210

[3] Department of Physics and Astronomy, University of Iowa, Iowa City, IA 52242

[4]Bergmann Messgeraete Entwicklung, Hagener Leite 24, 82418 Murnau, Germany

[*]Corresponding author: Jay Gupta, Gupta.208@osu.edu

## ABSTRACT

We report scanning tunneling microscopy studies of individual adatoms deposited on an InSb(110) surface. The adatoms can be reproducibly dropped off from the STM tip by voltage pulses, and impact tunneling into the surface by up to ~100x. The spatial extent and magnitude of the tunneling effect are widely tunable by imaging conditions such as bias voltage, set current and photoillumination. We attribute the effect to occupation of a (+/0) charge transition level, and switching of the associated adatom-induced band bending. The effect in STM topographic images is well reproduced by transport modeling of filling and emptying rates as a function of the tip position. STM atomic contrast and tunneling spectra are in good agreement with density functional theory calculations for In adatoms. The adatom ionization effect can extend to distances greater than 50 nm away, which we attribute to the low concentration and low binding energy of the residual donors in the undoped InSb crystal. These studies demonstrate how individual atoms can be used to sensitively control current flow in nanoscale devices.

# INTRODUCTION

As electronic devices miniaturize to the nanometer scale, their performance and reproducibility becomes increasingly impacted by single-impurity effects [1,2]. Nanoscale electronics, such as single atom transistors[3], are a realization of a new 'solotronics'[4], where control and manipulation of individual impurity atoms in a material determines the device performance. In this context, the scanning tunneling microscope (STM) has been used for deterministic placement of dopants [refs] and to tune dopant binding energies via control of the local electrostatic environment [ref] [5] STM-induced ionization is also a promising demonstration for control, and has now been reported in various systems including dopants in GaAs[7–9], silicon dangling bonds [14–18], C60[10], cobalt on graphene[11], iron on $Bi_2Se_3$ [12] and vacancies in ZnO[13].

Here we report a tip-induced ionization effect that reveals how individual adatoms can be used to tune the tunneling conductance of an InSb(110) surface by two orders of magnitude over relatively large areas. When the adatoms are in a positive charge state, adatom induced band bending leads to a high-conductance accumulation region that can extend for >> 100 nm$^2$ due to the very low binding energy of native donors in InSb. Under certain imaging conditions, the adatom can be switched to a neutral state by a tunneling electron from the STM tip, leading to a low-conductance depletion region appearing as a 'crater' feature in topographic STM images. The crater size can be tuned with bias voltage, set current, and photoillumination. The experimental results are confirmed by comparisons with DFT calculations and theoretical transport modeling. These studies provide insight into how single atoms can be harnessed in future nanoscale opto/electronic devices.

# METHODS

Experiments were conducted in a Createc LT-STM at 5K in ultrahigh vacuum ($< 10^{-10}$ mbar) using electrochemically etched Pt/Ir tips with an apex diameter $< 500 nm$. The InSb sample was a commercial bulk wafer, nominally undoped, but with *n*-type conductivity verified with Hall effect measurements. Typical donor species (e.g., S, Se, Te) in InSb are thought to occur at a concentration in the $10^{13-14}/cm^3$ range, with a ~0.6 meV binding energy [29,30]. The sample was cleaved at ~100K in UHV, exposing the (110) surface, immediately before transfer into the STM. Tunneling spectroscopy was performed by adding a 10mV modulation to the sample bias. A lock-in detects the corresponding d*I*/d*V* (*V*) signal, which is related to the local density of states (LDOS). To tune band bending via the surface photovoltage (SPV) effect[19], light from a home-built laser (1550nm, 125 fs pulses at a 70 MHz repetition rate) was aligned to a side viewport on the STM chamber. An average laser power of 1.5 mW was directed at glancing incidence onto the sample with no additional focusing. The illuminated area on the sample was relatively large (~ 0.5 cm$^2$), so that careful alignment of laser and tip was not necessary.

All plane wave DFT calculations were performed using the projector augmented wave pseudopotentials provided in the Vienna ab initio simulation package (VASP).[31–33] Further details of the calculations are provided in the Supporting Information, but here we note that the bulk InSb band gap with Perdew-Burke-Ernzerhof (PBE) [31] is more accurate than the hybrid HSE06 functional[32] value. Based on this result we chose the PBE functional for all InSb(110) surface calculations reported in the paper. Surface relaxation was performed with the lateral dimensions fixed to the bulk PBE lattice constant, a force criteria of 0.03 eV/Å, a plane wave cutoff of 400 eV and Γ-centered 4×4×2 k grids. We tested the four most likely adatom candidates (In, Sb, Pt, Ir) on the site indicated by the STM images. Bader charge analysis[34] was performed for each of the adatom candidates using the program from Henkelman and co-workers.[35–37] We obtain local density of states (LDOS) of the In adatom and bare InSb surface using the same settings (k-point grid, plane wave cutoff) as for the bare surface relaxation. Lastly, we simulated constant-height STM images of adatoms on the InSb surface using the Tersoff-Hamman model[38] for both empty and filled state conditions.

**RESULTS AND DISCUSSION**

Figure 1a shows an STM image of the clean InSb(110) surface under filled-state imaging conditions where the rows of Sb atoms are resolved[19], as are a few bright features which correspond to a low coverage of molecular adsorbates from the UHV chamber. Larger scale images of the surface (Fig. S1) show very few native defects such as surface/sub-surface dopants or atomic vacancies, which we attribute to the low defect concentration in our sample. During tip forming procedures (e.g., approach, voltage pulsing) that are typically used to sharpen STM tips, we were able to produce new point defects on the surface that we attribute to individual adatoms. For example, Figure 1b shows the same area as Figure 1a after a +8V pulse that dropped an adatom off the tip. This process can be repeated multiple times, resulting in atoms dropped from the tip one by one. We attribute these defects to individual adatoms based on their small apparent size and reproducible appearance. If the tip were depositing few-atom clusters, we would expect variations in number, size and shape of the features that were not observed in the experiment.

In prior work on GaAs, we found that the STM tip readily becomes terminated with the substrate material during standard tip forming procedures, and we could deposit Ga adatoms in a similar fashion. [23] We attribute the defect in Fig. 1b to an In adatom dropped off from the tip based on a comparison of our experimental data with DFT calculations. Discussed further in Supporting Information, we first determined an interstitial adsorption site from atomic resolution images where both In and Sb atomic rows could be identified (c.f., Fig. 1c). We then performed DFT calculations of the four most likely candidates (i.e., In, Sb, or Pt/Ir from the tip) on this high symmetry site. Bader charge analysis was used to estimate the charge state of the adatoms, and we find that of the four candidates, only $In_{ad}$ is positively charged, which agrees with the experimentally determined charge discussed below. The DFT-simulated STM image for interstitial $In_{ad}$ (Fig. 1d) is in reasonable qualitative agreement with experiment (Fig. 1c), reproducing the enhancement of contrast along the neighboring Sb row (similar to a previous study of $Fe_{ad}$ on InSb[21]), and the dark node centered on the atomic site. In further support of this assignment, we discuss below the good agreement between experimental and calculated local density of states for the In adatom, and include additional details in the Supporting Information.

We now discuss the charge state of the adatom and tip-induced ionization effects. Figure 2 shows that the adatom images as a protrusion at negative bias (Fig.2a), and under positive bias conditions as a dark 'crater' feature (Fig 2b). Linecuts of the apparent topography as indicated in these images are shown in Figure 2c. At negative bias, the apparent topography shows a gradual fall-off that extends for several nm away from the adatom site, much further than the atomic states associated with the adatom. This is commonly observed for charged species on III-V (110) surfaces[24] and we attribute the change in topography to adatom-induced band bending (AIBB)[23] of a positive point charge. To understand this connection, we note that the tunneling probability (and thus apparent height in STM images) depends on the total band bending (TBB) at the surface underneath the tip. In our experiments, TBB is the sum of several contributions as given by Eq. 1:

$$TBB = TIBB(V,z) + AIBB(q,r) + \sum_{i=1}^{N} DIBB_i(q_i, \vec{r}_i) + SPV. \qquad (1)$$

Here, tip induced band bending (TIBB) is a function of bias voltage $V$ and tip height $z$, and switches sign at $V = V_{FB}$, where $V_{FB}$ is the flat band voltage. AIBB is a function of the adatom charge state $q$, and how far the adatom is from the tunneling apex, $r$; defect induced band bending (DIBB) represents contributions from the local environment (neglected here due to the low defect concentration in our samples), and SPV is the surface photovoltage contribution if the sample is being illuminated. The separation of photo-excited carriers in local electrostatic fields always acts to decrease the magnitude of TBB, regardless of its sign,[24] and the underlying dynamics can be spatially resolved near individual impurities.[6] Though it is difficult to determine TIBB directly, we performed a number of simulations using the SEMITIP Poisson solver from Feenstra[25] to estimate TIBB using typical values of tip sharpness, work function and bias voltage, which is discussed further in the Supporting Information. Based on the electron affinity of InSb and the estimated work function for the PtIr tips, we expect the flat band condition to be in the range $-0.2\,V < V_{FB} < 0.4\,V$ in our experiments.

The tunneling diagrams in Fig. 2d,e describe how the image contrast in Figs. 2a,b reflects the adatom charge state. Under the conditions in Fig. 2a (V < V$_{FB}$), we expect $TIBB < 0$, which should cause an accumulation of conduction electrons under the tip in our *n*-type sample. This accumulation region moves in concert with the tip, and while tunneling of these electrons into the tip's empty states determines the absolute tip-sample distance, it is largely invisible in the experimental images. The bright contrast near the adatom is consistent with a positive adatom charge state; $AIBB < 0$ increases accumulation near the adatom (green bands and arrow in Fig. 2d) and the tip moves further from the surface in the constant-current image. The gradual fall-off in Fig. 2c is consistent with a screened Coulomb potential [17], with a screening length of $\sim 3.5\,nm$ which depends on the imaging conditions through TIBB (*V*, *z*) (c.f., Fig. S2).

The crater feature at positive voltage (Fig. 2b), shows an abrupt change in apparent height of 150 pm, indicating more than a tenfold reduction in tunneling conductance. The Sb rows are observed within the crater feature and appear continuous with the rows outside of the crater, indicating this is an electronic effect rather than a true topographic feature. The bottom of the crater is relatively flat (Fig. 2c, blue curve), and there is no longer-range fall off in apparent height near the adatom position, suggesting that the adatom is now in a neutral charge state. Under the conditions of Fig. 2b (i.e., V > V$_{FB}$), $TIBB > 0$, which means that tunneling electrons from the tip must cross a depletion barrier to get to the InSb conduction band (Fig.2e, red arrow). This depletion region follows the tip as it is scanned, and is not readily apparent in STM imaging. Within a threshold distance, the STM tip switches the adatom to a neutral charge state. The abrupt increase in TBB when AIBB → 0 leads to an increased depletion barrier near the tip (green curve in Fig. 2d), so that the tip must get significantly closer to the surface to maintain a constant current, accounting for the apparent depth of the crater feature.

The abrupt change in charge state as the tip approaches the adatom is suggestive of a tip-induced ionization effect. Recently, disk- or ring-like features in STM images of impurities in other III-V semiconductors have been attributed to the ionization of individual dopants at a threshold value of TIBB that brings a dopant charge transition level into resonance with the host conduction or valence bands [7,9,26]. The size and shape of these features depend on factors which affect TIBB, including bias voltage, set current and the tip apex structure.

In Fig.3 we demonstrate a similar tunability of the crater feature. In our experiment, however an alternate ionization mechanism is necessary to consistently explain all the data. Compared to the crater under typical imaging conditions (Fig. 3a), we observed a significant reduction in size by three methods: (*i*) illuminating the sample with light (Fig. 3c), (*ii*) reducing the set tunneling current (Fig.3d), and (*iii*) increasing the bias voltage (Fig. 3b). Under illumination, SPV counteracts TIBB, so that the tip must move closer to the adatom to achieve the same threshold condition. Similarly, a reduced set current (Fig. 3d) moves the tip further from the surface and thus reduces TIBB, which shifts the threshold condition closer to the adatom as well.

While both of these observations are qualitatively consistent with a resonance TIBB ionization effect, the sensitivity of the crater to such small changes in set current and the *reduction* of the crater size with higher positive voltage (Fig. 3b) suggests an alternative mechanism is needed. Because we expect TIBB to further increase in magnitude with increasing voltage from +0.5V (Fig. 3a) to +0.9V (Fig. 3b), the threshold condition should be met at a larger tip-adatom distance (i.e., larger crater), in contrast to the observation. In the Supporting Information, we considered other possible scenarios given our uncertainties in $V_{FB}$ and other TIBB parameters. Our conclusion is that there is no resonance TIBB scenario that explains the trends in crater size with *V, I* while also being consistent with our assignment of a positive charge state when the tip is far away from the adatom, and a neutral charge state when the tip is brought nearby.

Following an approach developed for STM tip-induced ionization of dangling bonds in silicon[13,14,16], we propose a competing rate model and find good qualitative agreement between our observations and theoretical transport modeling (Fig. 4). Here, $TIBB > 0$ brings the adatom's (+/0) charge transition level into a double-barrier configuration, resembling a quantum dot coupled to two leads[15]. Our model includes one channel for the tunneling current characterized by a filling rate from the tip, $\Gamma_f$, and an emptying rate into the bulk, $\Gamma_e$. Once the adatom state is filled by an electron from the tip tunneling through the vacuum barrier, it must escape into the bulk by tunneling through a surface depletion barrier, with a height and width that depend on TIBB. The time-averaged occupation of the state is given by $N = \frac{\Gamma_f}{(\Gamma_f + \Gamma_e)}$ and reaches a value of ½ when $\Gamma_f = \Gamma_e$. Direct tunneling into the InSb surface is considered as a second tunneling channel, with rates $\Gamma_0$, $\Gamma_1$ that depend on the adatom charge state (neutral and positive respectively) via AIBB.

Of these four rates, we can estimate $\Gamma_0$, $\Gamma_1$ from tunneling into InSb with the tip positioned inside / outside the crater region, and we calculate $\Gamma_f(r)$ using a Tersoff-Hamann approximation treating the terminating tip atom as a spherically symmetric wavefunction, and the adatom wavefunction with separate vertical ($z_0$) and lateral ($r_0$) decay constants. The filling/emptying rates are difficult to directly estimate from our STM data because they are sensitive to the measurement conditions and can easily lie outside the limited instrumental bandwidth (~ 1 kHz), but we note that 'telegraph' noise is observed in STM images under certain conditions that is suggestive of these dynamics.

Figure 4 shows a plot of the apparent height at constant current from this calculation as a function of the separation between tip and adatom. As shown in the figure, the apparent height transitions from a low to high value as the tip moves away from the adatom. The transition reflects the threshold separation where $\Gamma_f \sim \Gamma_e$. For $\Gamma_{e,f} \sim$ 10 kHz (black curve in Fig. 4b), our calculations predict that the transition should occur at a separation of ~ 5nm, which is in qualitative agreement with the experimental image in Fig. 3a. Applying this model now to the trends in Fig. 3, we note that under photoillumination, the width of the depletion barrier decreases, which increases $\Gamma_e$ but has no direct effect on $\Gamma_f$. The tip must therefore approach closer to the adatom to balance the filling and emptying rates, consistent with the reduced crater size in Fig. 3c and Fig. 4b (green curve). Reduction of set current directly decreases $\Gamma_f$, so that the tip must move closer to the adatom to increase the depletion barrier width and decrease $\Gamma_e$, consistent with Fig. 3d. Higher positive voltage affects both rates with opposite effects on crater size: $\Gamma_e$ decreases because of a wider depletion barrier, while $\Gamma_f$ decreases because tunneling electrons from the tip have a higher vacuum barrier into the adatom state (neglecting the possibility of inelastic electron tunneling). Reduction in $\Gamma_f$ decreases the crater size while a reduction in $\Gamma_e$ increases the crater size. The data in Fig. 3b are consistent with a net reduction in the $\Gamma_f/\Gamma_e$ ratio at higher positive voltage: the tip must approach the adatom to increase $\Gamma_f$ and achieve the threshold condition, resulting in a smaller crater. Further discussion of the model presented in Fig. 4 can be found in Supporting Information.

To identify the adatom charge transition level responsible for the ionization crater, we compared tunneling spectroscopy data with the tip positioned on pristine InSb (Fig. 5a, black) and the adatom (Fig. 5a, red). On InSb, there is a minimum in the measured dI/dV signal in the range ± 200 mV that roughly reflects the InSb band gap of 235 meV. A sharp rise at +200 mV is attributed to the InSb conduction band, and a more gradual rise below -200 mV reflects the valence band. Tunneling spectroscopy on the adatom (Fig. 5a, red) indicates enhanced conductance below -400 mV consistent with increased accumulation due to $AIBB < 0$, and a pronounced suppression of the conductance above +200 mV, consistent with tunneling through a depletion region when $AIBB = 0$. We interpret the peak at +220 mV as the (+/0) charge transition level, indicated with red arrow. The calculated local density of states (LDOS) from our DFT calculations show qualitatively similar behavior. The InSb(110) surface DOS (Fig. 5b, black) shows a minimum in the density of states around 0V. Importantly, in the In adatom density of states (Fig. 5b, red), a distinct peak appears near +250mV, which is in reasonable agreement with our tunneling spectroscopy results and is not present in the calculated DOS of the surface indium sites (Fig. 5b, blue).

Somewhat counterintuitively, the spatial extent of the crater feature sets a *lower* limit for the length scale over which the adatoms influence the tunneling conductivity of the InSb(110) surface. Because the crater corresponds to the region where $AIBB = 0$, the apparent height of the surface *outside* the crater must reflect $AIBB \neq 0$, or there would be no change in contrast due to the adatom charge state. To estimate the limit of the spatial extent, we show in Figure 6 a series of images as a function of voltage with a few adatoms deposited from the tip. With a modest decrease in bias voltage, the crater features eventually grow to exceed the entire scan area, indicating that individual adatoms influence the surface conduction over distances greater than 50 nm. We attribute the additional suppression of conductance on the left in Fig. 6c to filling from a point on the tip distinct from the tunneling apex. Although these points are further from the surface and don't contribute much to $\Gamma_0, \Gamma_1$, they can be closer to the adatom and have higher $\Gamma_f$. This behavior further distinguishes ionization in the competing rates model [15–17,28] from the resonant TIBB models [7,9,26].

We attribute the remarkable spatial extent of the adatom-tunable conductance to the low concentration and binding energy ($\sim 0.6\ meV$) of the residual donors in the undoped InSb crystal.[29] This ionization energy is comparable to $k_B T \sim 0.4\ meV$ at 5K, which suggests that a sizeable fraction of the donors can be readily ionized thermally or by the electrostatic potentials of individual adatoms. Indeed, the Coulomb potential for a $+e$ adatom falls below 0.6 meV at a distance of $\sim 140$ nm, assuming an effective static dielectric constant of $\varepsilon = 16.8$ for bulk InSb. This sensitivity distinguishes the ionization effect observed here from previous studies and suggests that undoped semiconductors with a low concentration of very shallow dopants may be well suited for future solotronic devices.

**CONCLUSIONS**

In this work we demonstrated tunability of the conductance of the InSb(110) surface by individual indium adatoms. Our experimental data are found to be in good agreement with DFT calculations and a transport model considering adatom and bulk tunneling channels. Similar effects to those described above were observed at step-edges and near larger nanoclusters that were occasionally deposited onto the surface during tip-forming, suggesting this is a more general phenomenon that is not specific to the particular adatom species. STM-based pulse methods as demonstrated recently for the Si dangling bonds [17,28] would allow for direct correlation of tunneling rates with the local environment, thus providing a microscopic picture of electronic transport.

**ACKNOWLEDGEMENTS**

characterized the laser used for SPV measurements. All authors reviewed the manuscript and contributed to the writing of the text.

We acknowledge support from the Department of Energy under grant # DE-SC0016379. Laser construction and characterization by (E.L., F.B., K.W., E.C.) was funded by a grant from AFOSR (Grant # FA9550-12-1-0454). E.C and K.W acknowledges support from AFOSR grant FA9550-16-1-0069, and AFRL grant FA9451-14-1-0351. AA and DK were supported by NSF # 1809837 and acknowledge the Ohio Supercomputing Center for providing computational resources.

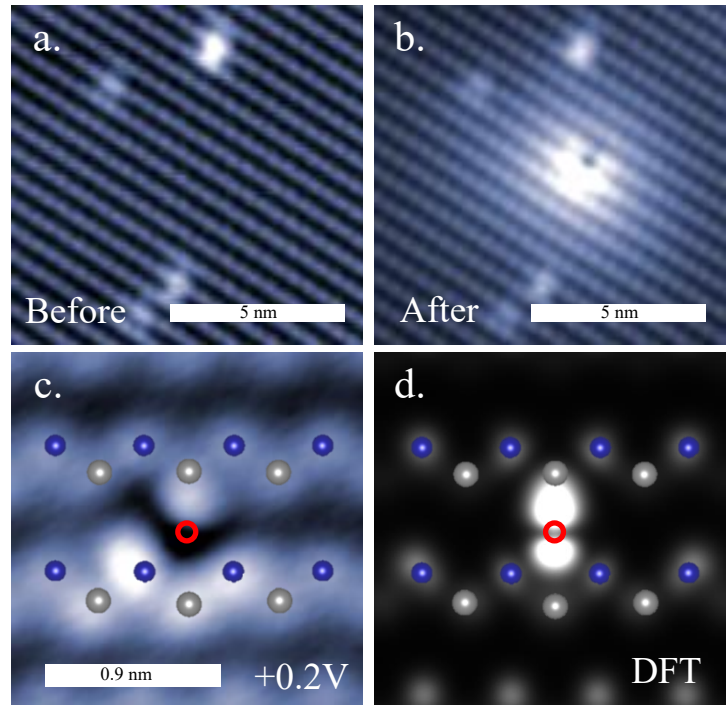

**Figure 1. Adatom Deposition and binding site assignment.** STM topographic images (a) before and (b) after a voltage pulse deposited an adatom to the InSb(110) surface (V = -0.5V, I = 0.22nA). (c) magnified view of adatom with InSb lattice overlay, blue atoms are Sb, grey are In, red circle is adatom site inferred from DFT (V = +0.2V, I = 0.22nA). (d) DFT simulated STM image. Red circle indicates the adatom position.

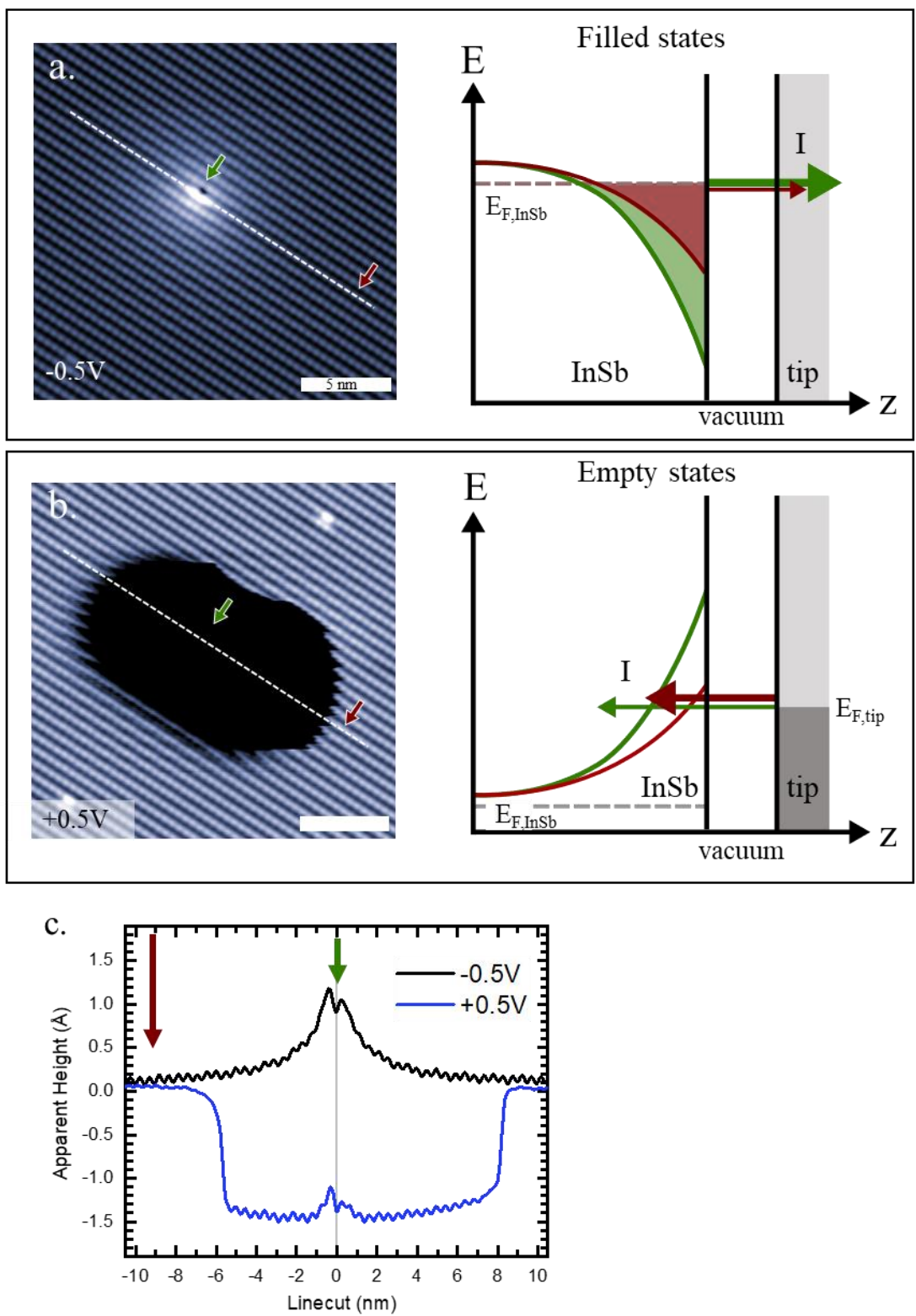

**Figure 2. STM-induced switching of adatom charge state** (a) Adatom under negative bias conditions (filled states) images as a protrusion (V = - 0.5V, I = 0.22nA). Accompanying tunneling diagram shows tunneling of electrons accumulated under the tip (solid regions). As the tip approaches the adatom (green), the additional accumulation from AIBB < 0 increases the tunneling probability compared to far from the adatom (red) (b) A dark 'crater' appears under positive bias imaging (empty states) (V = +0.5V, I = 0.22nA). When the tip is far from the adatom (red arrow), AIBB<0 but TBB > 0 and electrons tunnel from the tip into the InSb conduction band as shown by the red arrow in the accompanying tunneling diagram. Once a tunneling electron occupies the (+/0) adatom level, AIBB = 0 and increased TBB leads to a depletion barrier for tunneling electrons (green arrow). (c) Topography linecuts indicated by the dotted lines in (a,b) showing a 1.5 Å change in apparent height associated with the crater feature.

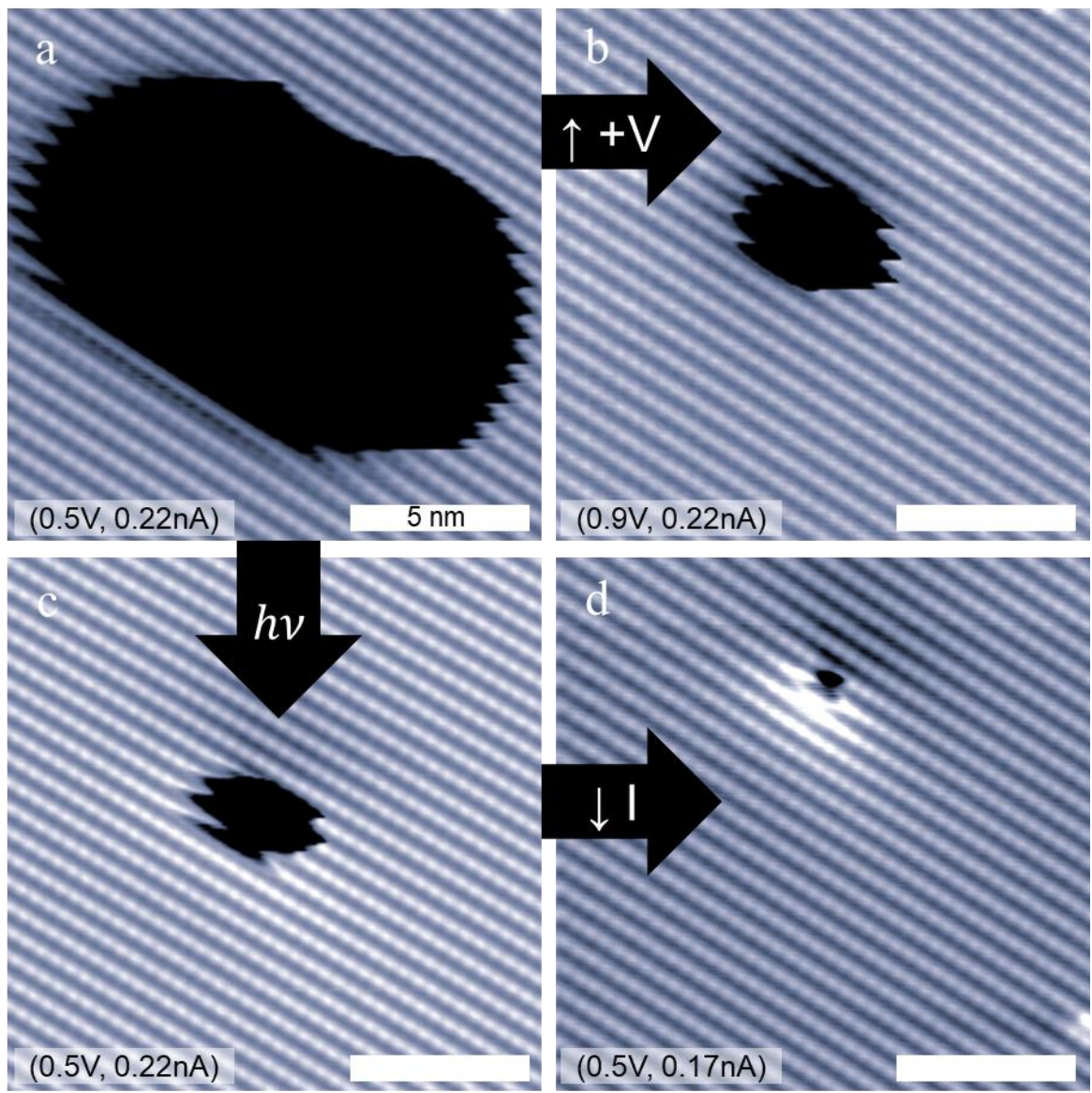

**Figure 3. Tuning of crater effect.** (a) STM image showing a crater feature under typical imaging conditions (+0.5V, 0.22nA). The crater size decreases (i.e., the tip must get closer to maintain occupation of the (+/0) level) by: (b) increased voltage (+0.9V, 0.22nA) or (c) photoillumination (+0.5V, 0.22nA) and (d) subsequent decrease in set current (+0.5V, 0.17nA). All scale bars are 5nm

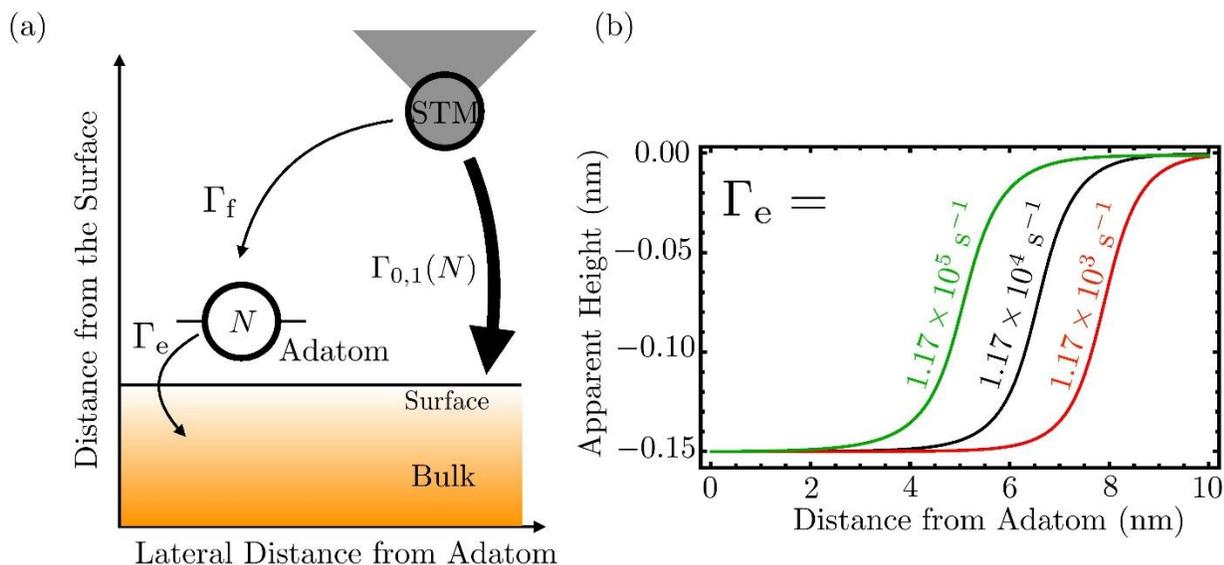

**Figure 4. Rates Modeling.** (a) Schematic showing tunneling channels through the adatom level and directly into the bulk. The total current is dominated by tunneling directly into the bulk as indicated by the weight of the arrows. Occupation of the (+/0) adatom level determines the magnitude of the bulk tunneling rate ($\Gamma_0$ or $\Gamma_1$) and is governed by the filling rate, $\Gamma_f$, and emptying rate, $\Gamma_e$. (b) Calculated apparent height based on the two-channel electrical transport model plotted for different values of $\Gamma_e$. For kHz-scale emptying rates, a change in apparent height occurs with the tip several nm away, qualitatively consistent with experiment.

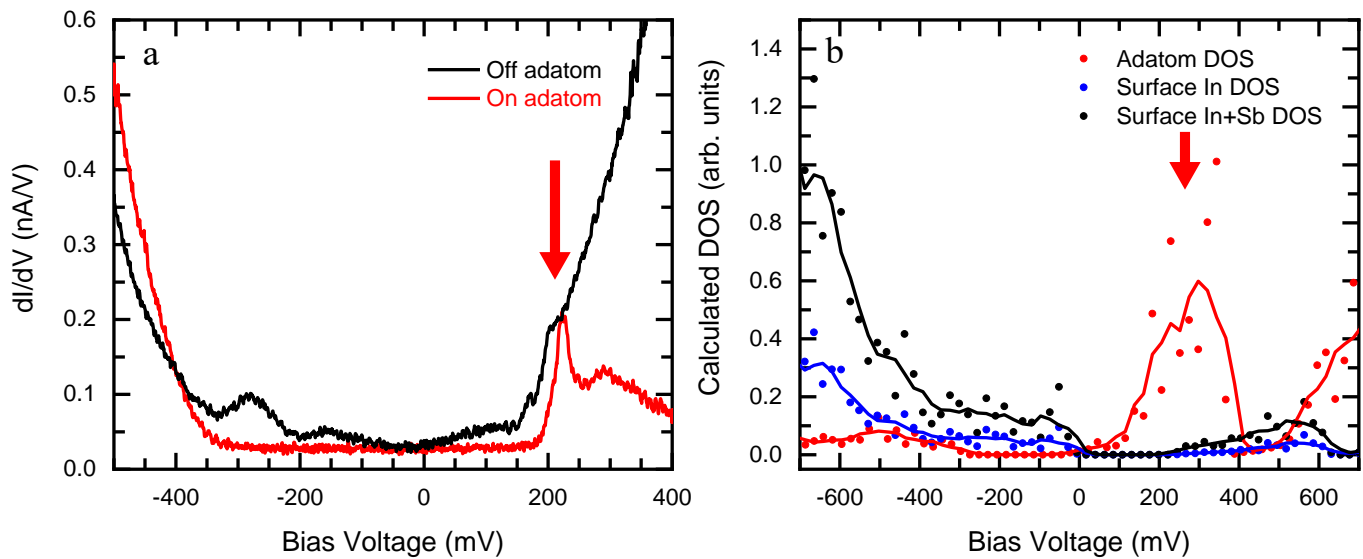

**Figure 5. Identification of the adatom charge transition level** (a) dI/dV spectroscopy on (red) and off (black) adatom showing a peak at 250mV (red arrow) attributed to the (+/0) charge transition level of the adatom. (b) DFT-calculated LDOS for the indium adatom (red), as well as surface In atoms (blue) and combined In and Sb atoms (black). Solid lines are smoothed curves serving as a guide to the eye. Red arrows indicate the (+/0) charge transition level of the adatom.

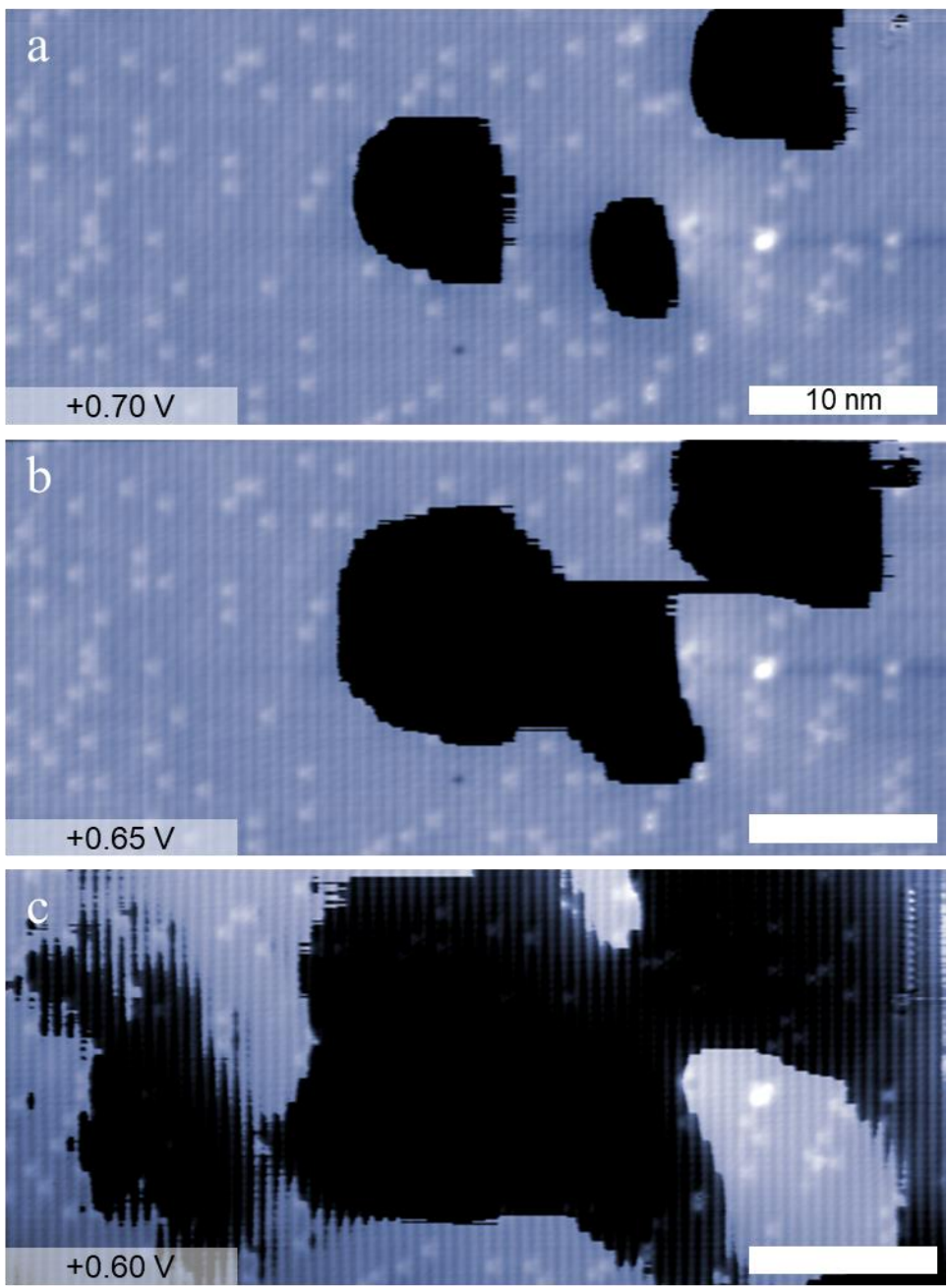

**Figure 6. Crater expansion.** (a) three crater features localized to their respective adatoms. (b) With a modest reduction in bias voltage, craters expand and overlap. (c) further reduction shows more overlap and development of an 'auxiliary crater' attributed to tunneling electrons into the adatom from higher up on the tip. (I = 0.22nA) Scale bars are 10nm.